\documentclass[aps,prl,twocolumn,showpacs,superscriptaddress,
]{revtex4-1}
\usepackage{mathrsfs}
\usepackage{graphicx}

\usepackage{epstopdf}
\usepackage{bm}        
\usepackage{amssymb}   
\usepackage{amsmath}
\usepackage{color}
\usepackage[toc,page]{appendix}
\usepackage{mathtools}

\def\bea{\begin{eqnarray}}
\def\eea{\end{eqnarray}}

\def\p{\partial}

\begin{document}


\title{Modeling of Ribosome Dynamics on a ds-mRNA under an External Load }

\author{Bahareh Shakiba}
\email{bahaareh.shakiba@gmail.com}
\affiliation{Department of Physics, Institute for Advanced Studies in
Basic Sciences (IASBS), Zanjan 45137-66731, Iran}

\author{Maryam Dayeri}
\affiliation{Department of Biological Sciences, Institute for Advanced Studies in
Basic Sciences (IASBS), Zanjan 45137-66731, Iran}

\author{Farshid Mohammad-Rafiee}
\email{farshid@iasbs.ac.ir}
\affiliation{Department of Physics, Institute for Advanced Studies in
Basic Sciences (IASBS), Zanjan 45137-66731, Iran}
\affiliation{Department of Biological Sciences, Institute for Advanced Studies in
Basic Sciences (IASBS), Zanjan 45137-66731, Iran}

\date{\today}

\begin{abstract}
Protein molecules in cells are synthesized by macromolecular machines called ribosomes. According to recent experimental data, we reduce the complexity of the ribosome and propose a model to express its activity in six main states. Using our model, we study the translation rate in different biological relevant situations in the presence of external force, and translation through the RNA double stranded region in the  absence or presence of the external force. In the present study, we give a quantitative theory for translation rate and show that the ribosome behaves more like a Brownian Ratchet motor. Our findings could shed some light on understanding behaviors of the ribosome in biological conditions. 
\end{abstract}

\maketitle

\section*{I. Introduction}

One of the most important processes in living cells is the {\it translation} in which the ribosome catalyses the synthesis of proteins from aminoacyl transfer RNA (tRNA), using messenger RNA (mRNA) as the template. In the ``initiation'' state, the two subunits of the ribosome join together near the 5' end of the mRNA. Following this state, the ribosome slides along the mRNA and translates its genetic information into an amino acid chain. In this process the transfer RNAs (tRNAs) have been used as adaptors for adding a right amino acid to the end of the polypeptide chain. The ``elongation'' process is followed by the ``termination'' state at the stop codon, where the ribosome finishes the protein synthesizes and the two ribosomal subunits separate \cite{Cell,Ramakrishnan-2009}.

Crystallography and cryoelectron microscopy experiments have revealed the structure of the ribosome in the atomistic level \cite{Yusupov-2001,Harms-2001,Schuwirth-2005,Selmer-2006}. The mRNA lies in the cleft of the small ribosomal subunit in such a way that its codons may interact with the anticodons of tRNA in three distinct binding sites, called the ``A'', ``P'' and ``E'' sites. At the A-site an incoming tRNA that carrying the next amino acid, binds to the associated codon in mRNA. The polypeptide chain is attached to the tRNA, which is located at the P-site. The deacylated tRNA leaves the ribosome at the exit or ``E'' site.

\begin{figure}
\centering
\includegraphics[width=1.0\columnwidth]{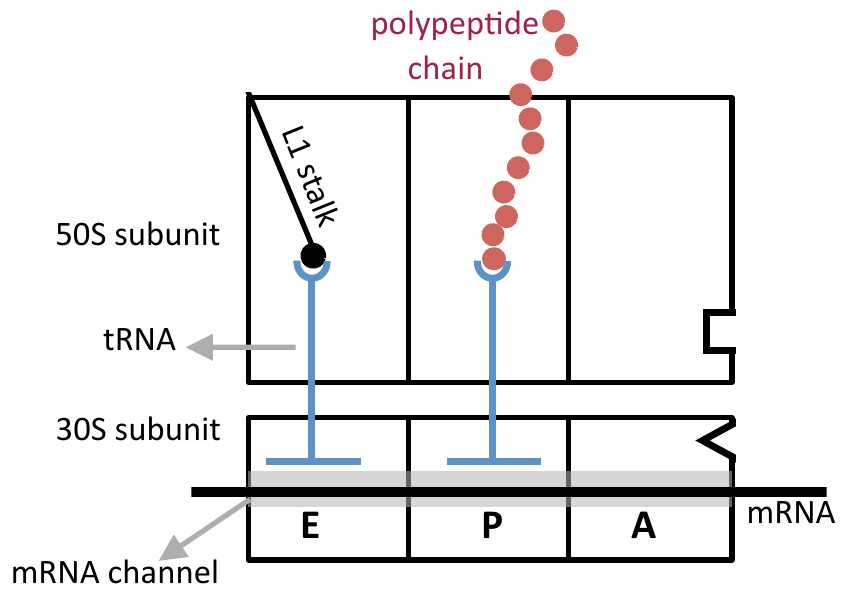}
\caption{(Color online) The Schematic picture of the ribosome in our model. Three main binding sites of the ribosome are shown with letters of E, P, and A. The mRNA channel is shown in grey, whereas the tRNAs are in blue. The aminoacids in the polypeptide chain are shown with the small red circles.  }
\label{fig-schematic-ribosome}
\end{figure}

We are investigating the generation of force during translocation with high-resolution optical traps. Additionally, we are also interested in how the ribosome uses force to overcome translational barriers such as secondary structure.

Since the ribosome is a huge complex catalytic machine with about 50 ribosomal proteins and several RNAs and also has several degrees of motional freedom, it is very difficult to monitor the translation process in experiments. 
However the possible mechanisms for ribosomal translocation have been developed in the recent years, taking different approaches that include molecular dynamics simulations \cite{Ishida-2014} and stochastic models \cite{Garai-2009, PingXie-2013, Bailey-2014}. In addition the overall sliding of the ribosome along the RNA have been examined using the normal mode analysis \cite{Chacon-2003, Stember-2009}. Furthermore, thanks to the optical tweezers techniques, now it is possible to study the force generation during the translation process, and the effect of the barriers like secondary structure of mRNA on ribosome translocation \cite{Bustamante-2008, Bustamante-2015}.
It is still a question how the chemical catalysis in the ribosome is coupled to its mechanical translocation. There are two basic schemes \cite{Wang-2002}: the Brownian ratchet (BR) and the power stroke (PS) mechanisms. The problem would be more complex when the ribosome encounters to a pseudo-knot of a folded RNA. There is another important question regarding the mechanism of unwrapping of the RNA double stranded region. The recent experiments study the ribosomal translocation through a hairpin structure of a mRNA. The results suggest that the hairpin can be unwound due to either the thermal fluctuations or the pushing past by the ribosome \cite{Bustamante-2011}. To address to mentioned questions, we attempt to propose a stochastic model to account for some observations for the movement of the ribosome on mRNA. According to the experimental observation, the elongation cycle of the ribosomal translation consists of many steps (more than 10 steps, see Appendix A for details) \cite{Wintermeyer-2004}. For the first step it is a good idea to reduce its complexity as much as possible and express its activity in a few main states for further investigations. The result of our simplification is shown in the schematic picture of a ribosome with its subunits and the mentioned sites in Fig. \ref{fig-schematic-ribosome}.
We describe the translation process with six key steps according to the recent experimental data (reviewed in \cite{Frank-2010}), as shown in Fig. \ref{fig-Model}.

\section*{II. Analysis}

At the beginning of the elongation cycle, ``state 0'', the A site is empty. The aa-tRNAs can bind to the vacant site A with a rate of $\omega_{01}$.  This bound aa-tRNA is in a turnery complex with an elongation factor EF-Tu and a GTP. If the bound aa-tRNA is matched to the mRNA codon at the A site, the transition from state 0 to sate 1 is occurred \cite{Blanchard-2004,Daviter-2006}. 
The GTP hydrolysis and the release of EF-Tu-GDP, promote a transition from state 1 to state 2, called the ``accommodation state''. In the accommodation state, the tRNA at the P site is joint to the tRNA at the A site through a peptidyl bond. In this state the tRNA of the A site is twisted and as a result the anticodon and the associated mRNA is displaced by 9 $\AA$ into the entry channel of the ribosome \cite{Noller-2002, Ramakrishnan-2009}.  It is worth mentioning that the polypeptide chain in this state is still attached to the tRNA at the P site. Through the state 2 to the state 3, the stress is released by transferring of the polypeptide chain to the tRNA of the A site, and the two tRNAs of the P and A sites are detached from each other \cite{Ramakrishnan-2009,Voorhees-2009}. The free tRNA at the P site has an affinity to the 50S E site, whereas at the same time the 50S P site has a specific interaction with the peptidyle tRNA. The mentioned interactions make the state 3 unstable and drive a ratchet like transition from the state 3 to the state 4 or ``the hybrid state'' \cite{Ramakrishnan-2009,Agirrezabala-2008}. We note that in the state 4, binding of an elongation factor, EF-G-GTP, stabilizes the hybrid state. The hydrolyses of GTP changes the conformation of the factor of EF-G and opens the mRNA channel. As the result, the 30S subunit is displaced by one codon toward to the downstream of the mRNA. At the same time both tRNAs that are bound to the mRNA, keep their position on the mRNA substrate (state 5) \cite{Valle-2003}. By dissociation of the tEF-G-GDP, the A site becomes empty and the two ribosomal subunits are aligned, whereas the ribosome goes from state 5 to state 0 \cite{Ramakrishnan-2009}. 

\begin{figure}
\centering
\includegraphics[width=1\columnwidth]{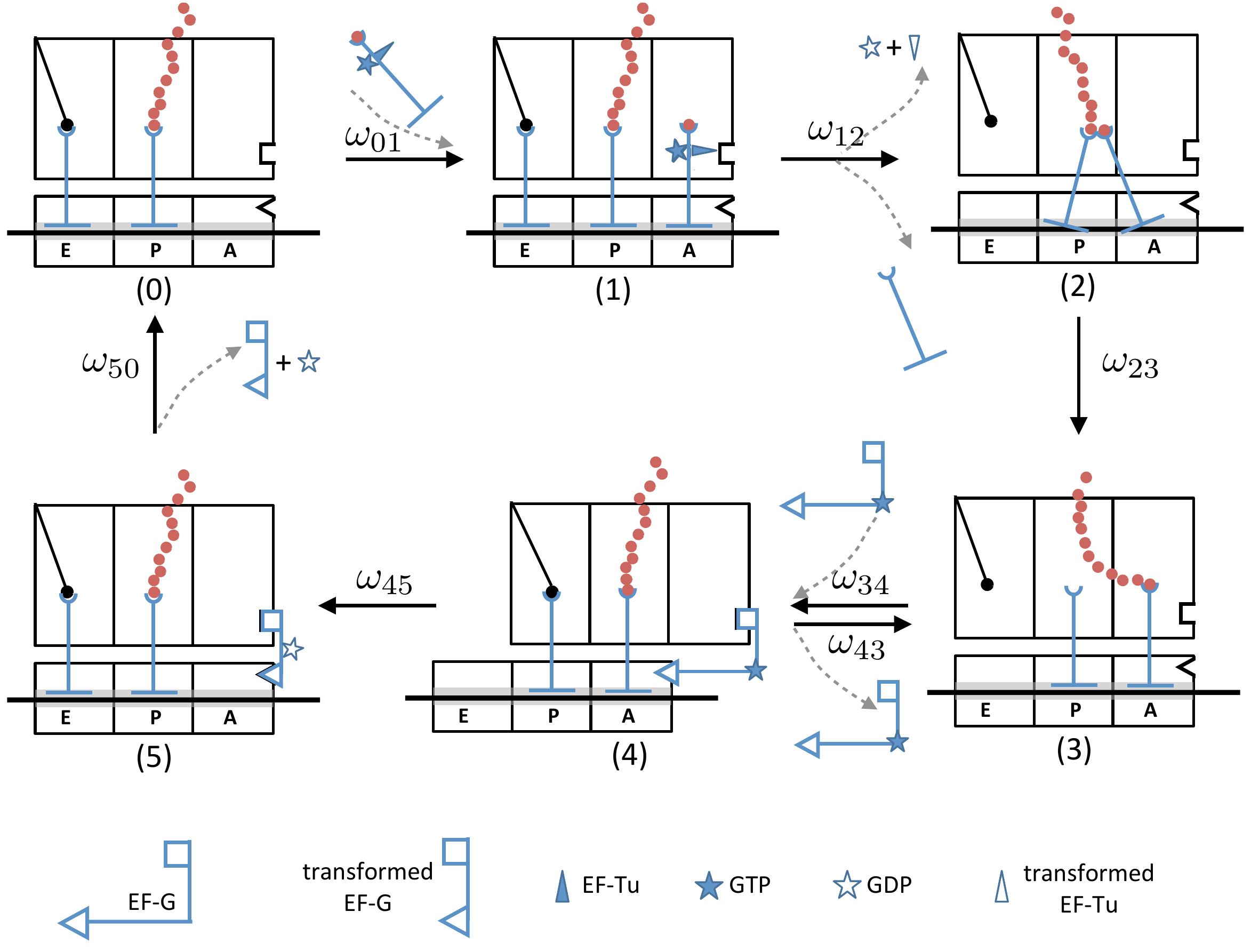}
\caption{(Color online) Different states of the translation process. 
Step $(0) \rightarrow (1)$: an EF-Tu-dependent aa-tRNA binds to the A-site of the ribosome. Step $(1) \rightarrow (2)$: the GTP hydrolyzes, the EF-Tu is deformed and released, and as a result a peptidyl bond is formed between the tRNAs of the A- and P-sites. $(2) \rightarrow (3)$: the polypeptide chain is transferred to the tRNA of the A-site. $(3) \rightleftharpoons (4)$: the elongation factor EF-G binds to the ribosome and promotes a ratchet like transition between state (3) and state (4). $(4) \rightarrow (5)$: the GTP hydrolysis changes the conformation of the factor EF-G and drives the unlocking of the mRNA channel and followed by mRNA movement. $(5) \rightarrow (0)$: the EF-G is released and the mRNA channel is relocked. The main elements of the translation process are shown in the last line.}
\label{fig-Model}
\end{figure}

To model the dynamics of the ribosome, consider the one-dimensional lattice with 6 different sites as described in Fig. \ref{fig-Model}. The ribosome can hop to neighboring sites on this lattice with some specific rates. The probability for the ribosome to be in the states $i=0-5$, at the position $x \equiv n a$ at time $t$ is denoted by $p_i(n,t)$, where $a$ is the length of one codon.
The $p_i(n,t)$ satisfies the master equation
\begin{subequations}
\bea
&&\p_t p_0(n,t) =  \omega_{50} p_5(n,t)- \omega_{01} p_0(n,t), \label{eq:master_a} \\ 
&& \p_t p_1(n,t) =   \omega_{01} p_0(n,t)  - \omega_{12} p_1(n,t),\\
&& \p_t p_2(n,t) =  \omega_{12} p_1(n,t) - \omega_{23} p_2(n,t),\\
&& \p_t p_3(n,t) =   \omega_{23} p_2(n,t) + \omega_{43} p_4(n,t) - \omega_{34} p_3(n,t), \\
&& \p_t p_4(n,t) =   \omega_{34} p_3(n,t) - \omega_{43} p_4(n,t) - \omega_{45} p_4(n,t), \\
&& \p_t p_5(n+1,t) =   \omega_{45} p_4(n,t) - \omega_{50} p_5(n+1,t), \label{eq:master_f}
\eea 
\end{subequations}
where $\omega_{ij}$ represents the rate of transition from state $i$ to neighboring state $j$. As we mentioned before, between every two states in our model, there are a few intermediate states that have been discussed in the Appendix A. Each rate in this model has been derived using the concept of net rate constants that will be discussed in the Appendix A. It is worth noting that at each translocation step, ribosome moves three nucleotides along the mRNA. In terms of these rates, one can find the mean translation velocity as (see Appendix B for details)
\bea
v = \frac{1 \; {\rm (codon)}}{ \frac{1}{\omega_{01}} + \frac{1}{\omega_{12}}  + \frac{1}{\omega_{23}} + \frac{1}{\omega_{34}} \left( 1 + \frac{\omega_{43}}{\omega_{45}} \right)  + \frac{1}{\omega_{45}} + \frac{1}{\omega_{50}} }. \label{eq:v-theory}
\eea

At each state, one can define a free energy energy, $G$, that has contributions of structural energy, $U$, and the chemical energy, $\mu$, as $G = U + \mu$. For example the external force as well as the concentration of the GTP can affect the free energy. In this paper for the sake of simplicity we ignore the effect of the mRNA sequence of nucleotides on the dynamics of the motor (i.e., no n-dependence of the rates). In order to understand the effects of the GTP concentration and the external force on the behavior of the ribosome, we use the idea that the possible reaction at each state proceeds through an activated state with a higher energy \cite{Howard,Rob-Philips}. Now one can write the rate transition of $\omega_{ij}$ and $\omega_{ji}$ as
\begin{subequations}
\bea
\omega_{ij} = k \, e^{-\beta \Delta G_{a,ij}}, \quad \Delta G_{a,ij} \equiv G_{a,ij} - G_i, \\
\omega_{ji} = k \, e^{-\beta \Delta G_{a,ji}}, \quad \Delta G_{a,ji} \equiv G_{a,ij} - G_j, 
\eea
\end{subequations}
where $G_{a,ij}$ denotes the activation free energy in the transition between two adjacent states of $i$ and $j$, and $k$ is the frequency factor for the mentioned transition. 

In the elongation state of the ribosomal movement, we have the hydrolization of the GTP:
$A.GTP \rightarrow A.GDP + Pi$, where $A$ denotes the elongation factors of EF-G or EF-TU as discussed above. It is plausible to consider that the external force only has effects only on $G$ as
\bea
G(F,[A.GTP]) &=& G_0 - F \Delta x, 
\eea
$F \Delta x$ shows the change in the free energy by the work of the motor against the external force, $F$. In the above equations, $G_0$ is the term that does not depend on the external force.

\section*{III. Results and Disucssion}

In the following we study the behavior of the ribosome in three biological relevant situations: (1) in the presence of external force, (2) translation through the RNAds (RNA double stranded) region in the absence of the external force, and (3) translation through the RNAds region in the presence of an external force.  Besides the analytical description of the problem, we perform a stochastic simulation of our model using the Gillespie algorithm \cite{Gillespie}.

\subsection*{Effect of External Force on the Ribosomal Translocation}

\begin{figure}
\centering
\includegraphics[width=1.0\columnwidth]{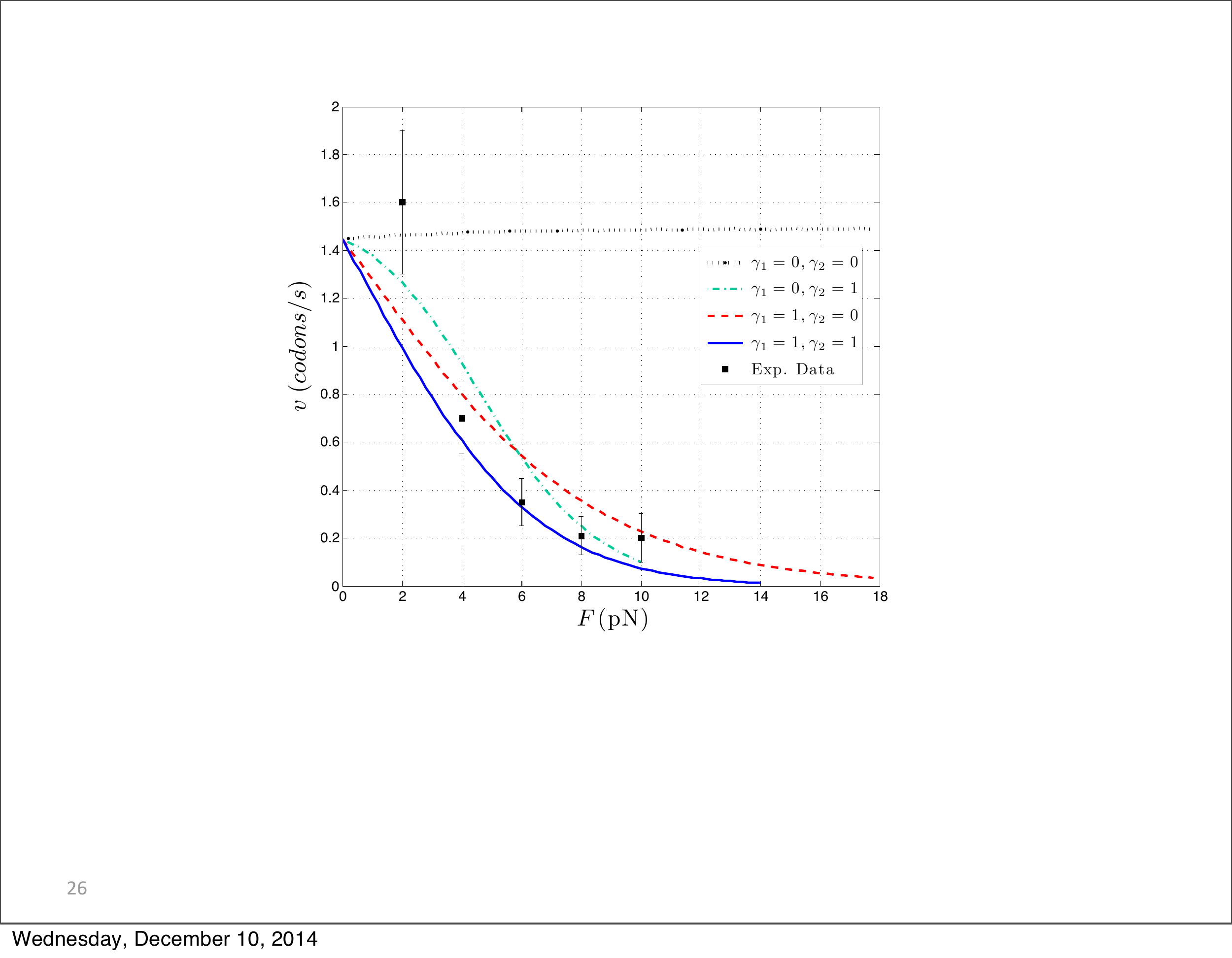}
\caption{(Color online) The translation velocity in terms of the external force for different mechanisms. The filled squares are the experimental data taken from Ref. \cite{Bustamante-2014}, in which the small subunit of the ribosome is fixed. The different lines are the simulation data corresponding to the different mechanisms. $\gamma_i = 0$ and $\gamma_i=1$ correspond to power stroke (PS) and brownian ratchet (BR) mechanisms, respectively. }
\label{fig:v-F}
\end{figure}

In the elongation cycle, in two states the ribosome displaced with respect to the mRNA. In the state 1, the GTP hydrolization drives the conformational change in the A-site tRNA, as mentioned above. This deformation pulls the mRNA into the mRNA channel by $9 \AA$. After that peptidyl transferring relaxes the stretched mRNA. Since there are some displacement in the mentioned processes, the external force can influence the transition rates correspondingly. Furthermore the transition between un-ratchet and hybrid state can be affected by the external force. It is worth mentioning that in the experiment it is possible to fix the large subunit or the small subunit. The way of fixing the ribosome in the experiment may affect the rates used in our model. After defining $\omega_{ij}^0 \equiv k \exp[-\beta \Delta G_{0, ij}]$, when the small subunit is fixed the rates are
\begin{subequations}
\bea
\omega_{12} &=& \omega_{12}^0 \, e^{- \gamma_1 \beta F a_1}, \label{eq:rate_F_small_1} \\
\omega_{23} &=& \omega_{23}^0 \, e^{+ \gamma_1 \beta F a_1}, \\
\omega_{45} &=& \omega_{45}^0 \, e^{-\gamma_2 \beta F a_2}, \label{eq:rate_F_small}
\eea 
\end{subequations}
where $a_1$ and $a_2$ are about $9 \AA$ and $21 \AA$, respectively and $0 \leq \gamma_{1,2} \leq 1$ are load distribution factors. It is known that $\gamma \simeq 1$ and $\gamma \simeq 0$ correspond to the``brownian ratchet'' and the ``power stroke'' mechanisms, respectively \cite{Howard}. It is worth to mentioning that we can write the similar equations for the case where the large subunit of the ribosome is fixed. For the transition rates, we use the parameters that are summarized in Table 1.

\begin{table}[b]
\centering
\caption{\label{tab:rates} The rate of transition from state $i$ to state $j$ used in our model for the movement of the ribosome. The rates are estimated using the concept of net rate constant and the values of Table \ref{tab:exp_rate} in Appendix A.}
\begin{ruledtabular}
\begin{tabular}{cccccccc}
parameter & $\omega_{01}^0$ & $\omega_{12}^0$ & $\omega_{23}^0$
& $\omega_{34}^0$ & $\omega_{43}^0$ & $\omega_{45}^0$ & $\omega_{50}^0$\\
\hline rates (${\rm s^{-1}}$) & 46 & 3 & 50 & 150 & 140 & 31 & 4 \\
\end{tabular}
\end{ruledtabular}
\end{table}

In Fig. \ref{fig:v-F}, the mean velocity of the translation is shown as a function of the external force, $F$, when the small subunit of the ribosome is fixed. The translation velocity is found using the rates of the Table 1 and as we discussed above, the force affects the rates according to Eqs. (\ref{eq:rate_F_small_1})-(\ref{eq:rate_F_small}). As can be seen in the figure, the translational velocity is decreased exponentially. We define a ``stall force'' for the motor where the velocity of the ribosome is reduced to less than 0.1 nt/s = 0.03 codons/s. The stall force used in our model is about $15$ pN, which is in a very good agreement with the recent experimental results \cite{Bustamante-2014}.

\subsection*{Effect of RNAds on the Ribosomal Translocation}

In the translational process, a 15-basepairs' (bp) long stretch of RNA is positioned in the mRNA channel between the A-site and the front side of the ribosome \cite{Seyedtaghi-2005}. Since the stretching modulus of the sugar-phosphate backbone is large enough, say $\sim 120 \, {\rm k_BT/nm^2}$ \cite{Zhang-2001}, we can consider this confined segment of RNA as a rigid rod. In the physiological condition, a single stranded RNA can be deformed into its secondary structure. Since the diameter of the mRNA channel is almost equal to the diameter of the single stranded RNA, $\sim 1 \, nm$, the front site of the mRNA channel prevents entering the double stranded segment of the RNAds. In order to translate through these RNAds regions, the ribosome should somehow unfold the RNA. In an active process, this happens either by applying a direct force by the ribosome \cite{Bustamante-2011} or by helicase activity of the small subunit of the ribosome \cite{Seyedtaghi-2005}. In a passive process the ribosome can be paused until the double stranded RNA unwinds due to thermal fluctuations \cite{Bustamante-2011}. We note that in principle modeling the unwinding of the double stranded RNA needs an understanding of the intramolecular interactions in the atomic length scales \cite{Stember-2007}.

In order to see the effect of the wound RNA on the translation rate, we should model the binding-unbinding of the bases of the RNA. The rate of binding and unbinding of two bases of the RNA are denoted by $k_{bind}$ and $k_{unbind}$, respectively. After introducing $\kappa$ as the ratio of these two rates and using the detailed balance condition, we have
\bea
\kappa \equiv \frac{k_{bind}}{k_{unbind}} = e^{- \beta \Delta G}, \label{eq:kappa}
\eea
where $\Delta G$ is the energy difference between the two mentioned states. Since the translation rate in the presence of pseudoknots depends on the overall unbinding of the RNA secondary structure, we study the translation rate in different values of the $k_{unbind}$. 

As mentioned before, in two states of the translational cycle, transitions $1 \rightarrow 2$ and  $4 \rightarrow 5$ , the small subunit of the ribosome moves along the mRNA by 1 bp and 3 bp, 
respectively. These transition rates can be influenced by the presence of the base pairs in the mRNA, which depends on the number of base pairs. Now we assume that a ribosome encounters a double stranded segment of the RNA. Let us discuss the transition of $1 \rightarrow 2$ and the effect of the wound mRNA. If the first base pair of the RNA is broken due to the thermal fluctuations (passive picture), the transition rate does not change and we will have $\omega_{12} = \omega_{12}^0$. But if the first base pair is still present, the ribosome should make it accessible and should break it, which costs energy and we have $\omega_{12} = \omega_{12}^0 \, \kappa$. For the transition of $4 \rightarrow 5$ the situation is very similar, the only difference is that the step length is now 3 bp. Since in the absence of double strands of the mRNA, the transition rate is denoted by $\omega_{45}^0$, in the presence of the wound area in the mRNA the transition rate would be changed to $\omega_{45} = \omega_{45}^0 \, \kappa^m$, where $m$ denotes the number of base pairs of the RNA in front of the enter channel that should be broken actively by the ribosome. So one can consider the mentioned transition rates as 
\bea
\omega_{12} &=& \omega_{12}^0 \, \kappa^m, \quad m = 0,1 \label{eq:omega12}\\
\omega_{45} &=& \omega_{45}^0 \, \kappa^m. \quad m = 0,1,2,3 \label{eq:omega45}
\eea

\begin{figure}[h]
\centering
\includegraphics[width=1\columnwidth]{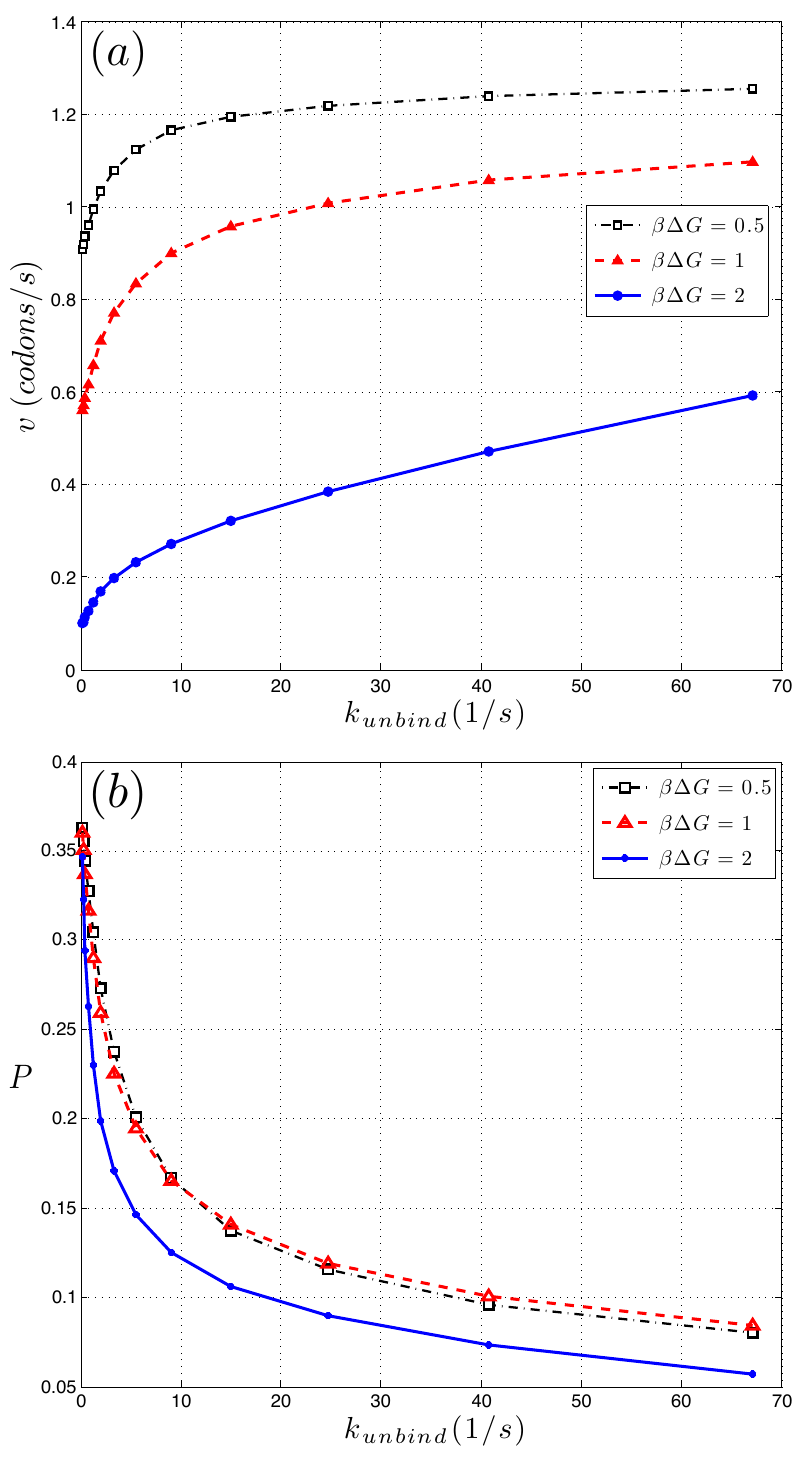}
\caption{(Color online) (a) The translation velocity and (b) $P$ as a function of un-binding rate of the base pairs of the dsRNA, respectively, for different values of $\Delta G$. $P$ denotes the fraction of actively opened to all opened base-pairs of dsRNA.}
\label{fig:v-k_unbind}
\end{figure}

The effect of the base-pair stability of the mRNA on the translational velocity is shown in Fig. \ref{fig:v-k_unbind}(a). When the unwrapping rate is high enough, the ribosome does not sense any barrier in front of it and translates easily the codes of the mRNA. As the base-pair of the RNA becomes stronger, the unwrapping rates decreases and therefore the velocity of the ribosome reduces. One may ask the following question: for a given value of the unwrapping rate, in how many cases the ribosome directly opens the base-pairs of the dsRNA and in how many cases these base-pairs are broken due to the thermal fluctuations? In order to answer this interesting question, we find the fraction of actively opened to the all opened base-pairs (actively or passively), $P$, in terms of the un-binding rate, as shown in the Fig. \ref{fig:v-k_unbind}(b). In Appendix C, the way of finding quantity $P$ has been explained. We see that when the $k_{unbind}$ is small, the ribosome has a crucial role in the breaking of the base-pair of the ds-mRNA. We can conclude that for small values of $k_{unbind}$, the ribosome directly breaks the base-pairs of the ds-mRNA by the probability around $0.3$.

\subsection*{Ribosome Translocation on a ds-mRNA under External Loads}

\begin{figure}
\centering
\includegraphics[width=1\columnwidth]{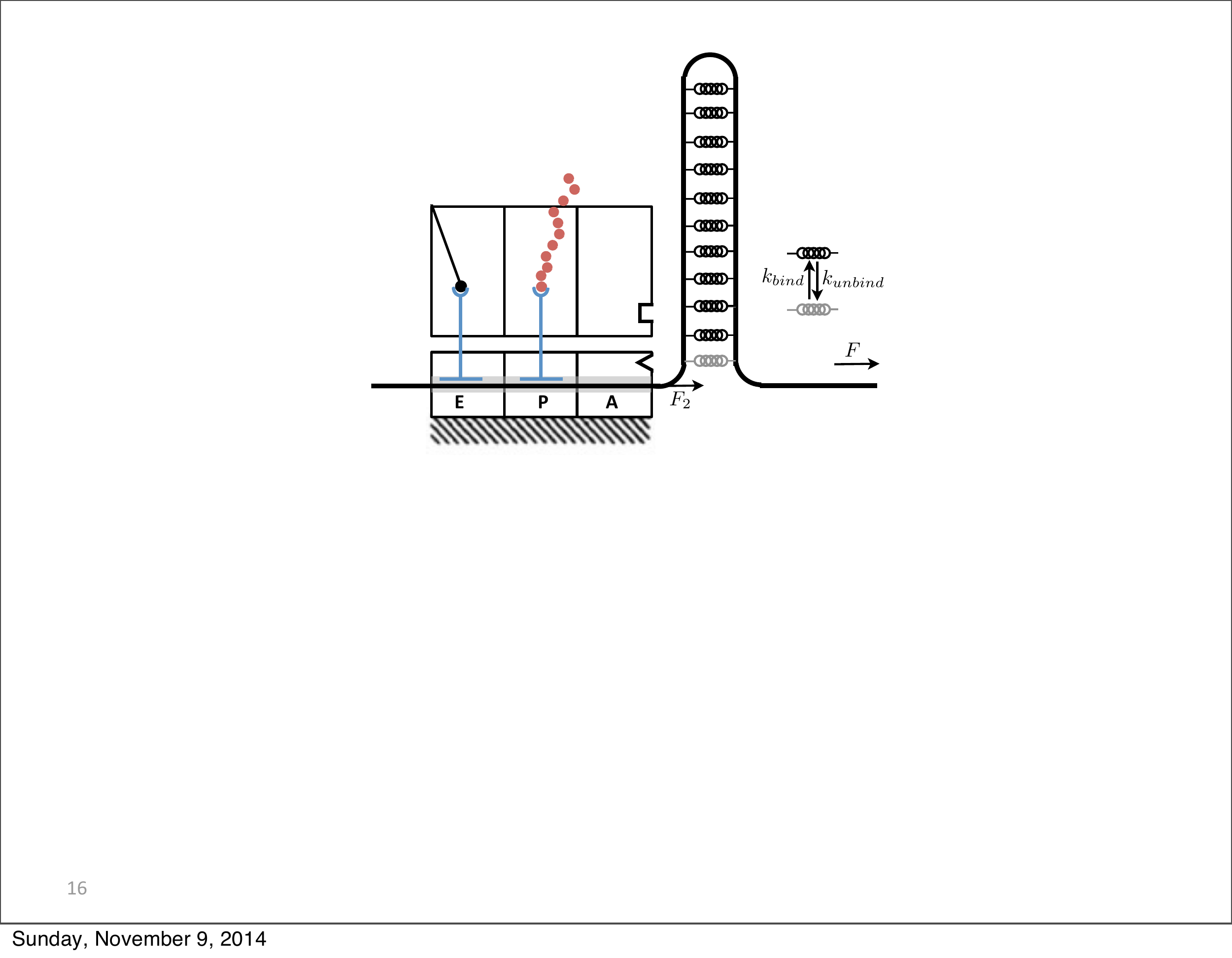}
\caption{The Schematic picture of the ribosome in the presence of a ds-mRNA and the external force, $F$. The force that is sensed by the ribosome is $F_2$. Two complimentary bases of the mRNA can be bound by the rate of $k_{bind}$ and the bond can be broken by the rate of $k_{unbind}$. In this figure, the small subunit of the ribosome is fixed. }
\label{fig:suggested-setup}
\end{figure}

Here we consider a situation at which the ribosome is facing the double stranded region while an external force is being applied to the other end of the mRNA. The schematic picture of the suggested setup is shown in Fig. \ref{fig:suggested-setup}. We can derive the corresponding changes in transition rates based on the discussions of the above sections. When the small subunit is fixed the rates are
\begin{subequations}
\bea
\omega_{12} &=& \kappa'^{m} \omega_{12}^0 \, e^{- \gamma_1 \beta F_2 a}, \\
\omega_{23} &=& \omega_{23}^0 \, e^{+ \gamma_1 \beta F_2 a}, \\
\omega_{45} &=&  \kappa'^{m} \omega_{45}^0 \, e^{- \gamma_2 \beta F_2 a}.
\eea 
\end{subequations}

\begin{figure}
\centering
\includegraphics[width=1\columnwidth]{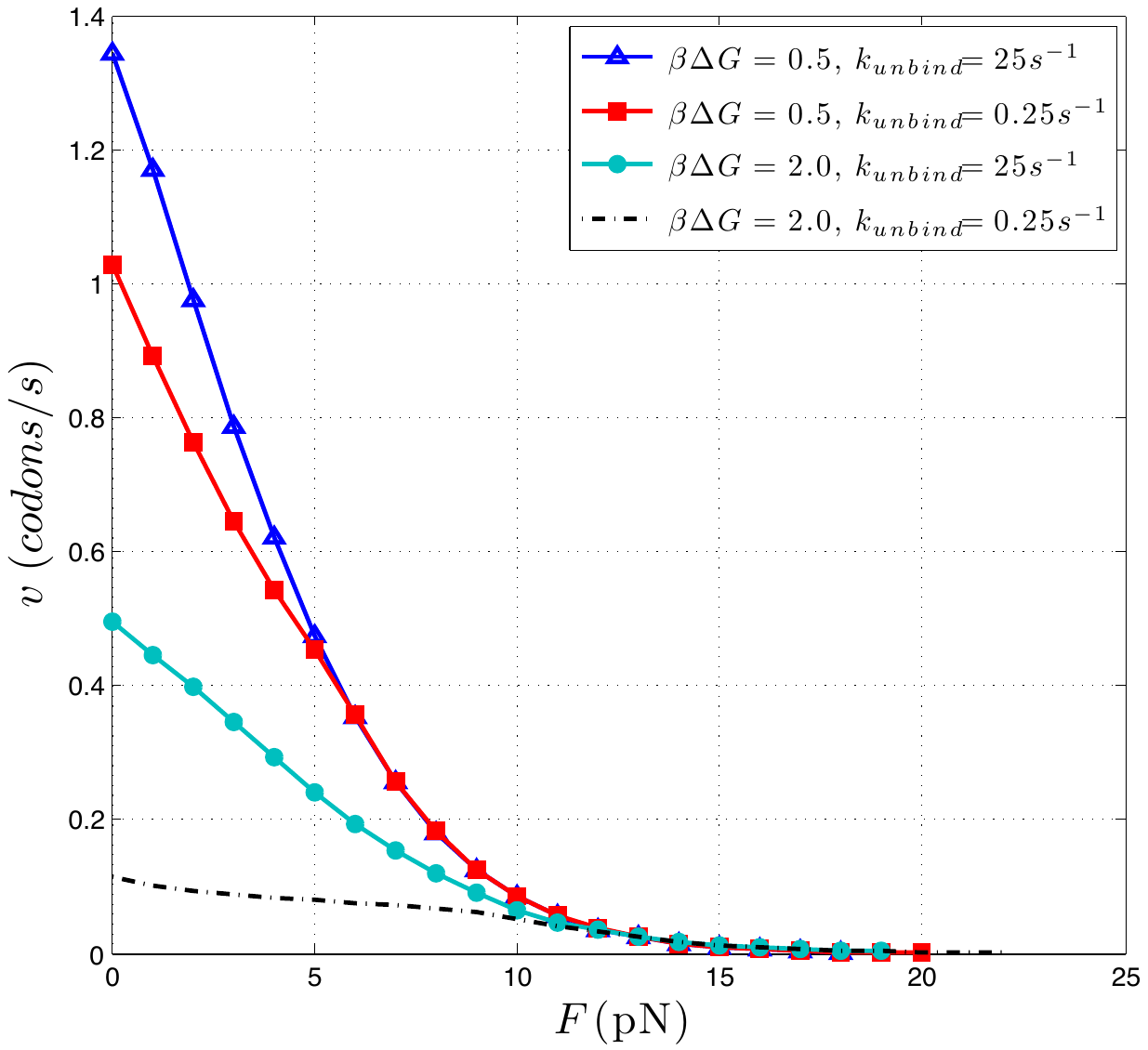}
\caption{(Color online) The translation rate as a function of $F$ for different values of $\Delta G$ and $k_{unbind}$.  }
\label{fig:v_F-dsmRNA}
\end{figure}

In the above equations we have used $\kappa' \equiv \kappa\, e^{\beta F x}$, where ``$F$'' is  the external applied force, for example is exerted by the optical tweezer. A contribution of the applied force breaks the bonds between mRNA bases, whereas a remaining contribution, $F_2$, affects the ribosome. 
We simply can model this process based on the molecular friction due to transient base-pairing of the mRNA bases. Let us assume that due to applying the external force, the base-pairs of the dsmRNA break with the velocity of $u$, number of codons per seconds. As we mentioned above in Eq. (\ref{eq:kappa}), the rate of binding and unbinding of two bases of the RNA are denoted by $k_{bind}$ and $k_{unbind}$, respectively and in principle both of them are functions of force. In terms of these rates, the mean velocity $u$ can be written as $u = \left(k_{unbind} - k_{bind} \right) b$, where $b$ is the length of one base step, say $b \sim 0.3$ codon. When the external force is zero, $k_{bind}(F=0) > k_{unbind}(F=0)$ and the mean velocity of the opening of the mRNA pseudo-knot is zero. The external force increases $k_{unbind}$ and decreases $k_{bind}$ and at some force, say $F^* = \frac{k_BT}{b} \ln \kappa^{-1} $, these two rates become equal, $k_{bind}(F=F^*) = k_{unbind}(F=F^*)$. Before this threshold, as $u$ is zero, the force that is affecting the ribosome is equal to $F$ and the ribosome moves like the situation discussed in the last section. When the applied force becomes larger than $F^*$, the ds-mRNA starts to open and we have a nonzero $u$ and the ribosome senses the force $F_2$. After denoting the phenomenological friction coefficient by $\mu$, we have $F_2 = F-\mu u$. We note that $\mu$ can be estimated in terms of $k_{unbind}$ and $\Delta G$, the energy difference between the ``bind'' and ``unbind'' states (see Eq. (\ref{eq:kappa})). There is an effective friction in the problem due to the transient crosslinks between two complimentary bases of the mRNA. If the stiffness of each base-pair is shown by $k \simeq \frac{\Delta G}{b^2}$, then the average force opposing the external force approximately is $  - \frac{\Delta G}{b^2 k_{unbind}} u$. The mean force that is acting on the ribosome can be estimated by
\bea
F_2 = \frac{k_{unbind}(F) - k_{bind}(F)}{k_{unbind}(F)} \, \frac{\Delta G}{b}.\label{eq:F2}
\eea

\begin{figure}
\centering
\includegraphics[width=1\columnwidth]{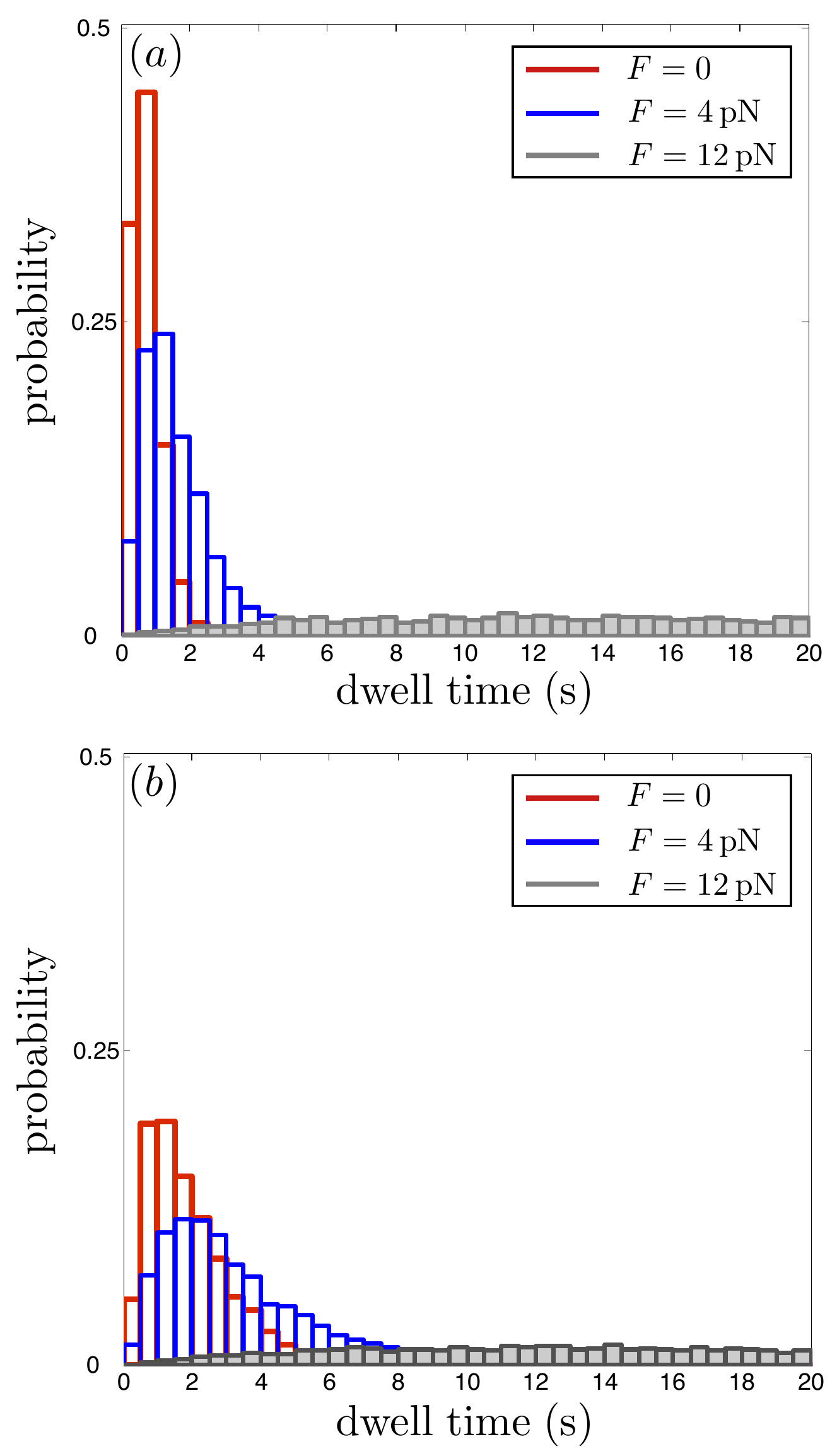}
\caption{(Color online) Histograms of dwell time for two situations. The red, blue, and grey histograms correspond to $F=0$, $F = 4$ pN, and $F = 12$ pN, respectively. Plot (a) corresponds to $ \beta \Delta G = 0.5, \; k_{unbind} = 25 \; {\rm s^{-1}} $, and plot (b) corresponds to $ \beta \Delta G = 2.0, \;  k_{unbind} = 25 \; {\rm s^{-1}} $. }
\label{fig:histogram-dt-F}
\end{figure}

In Fig. \ref{fig:v_F-dsmRNA} the behavior of the translation rate is shown as a function of the applied force, $F$, for representative values of $\Delta G $ and $k_{unbind}$. We note that in principle the unbinding rate of the G-C and A-T base pairs are different. As can be seen in the figure, for a given values of $\Delta G$, the translation velocity reduces as $k_{unbind}$ becomes smaller. In experiments, one can measure the time between successive steps in the translation process, which is called dwell time \cite{Bustamante-2008}. In Fig. \ref{fig:histogram-dt-F}, histograms of the dwell time for three forces for two different situations are shown. The Fig. \ref{fig:histogram-dt-F}(a) is corresponding to $\beta \Delta G = 0.5$ and $k_{unbind} = 25 \, s^{-1}$, and the Fig. \ref{fig:histogram-dt-F}(b) is corresponding to $\beta \Delta G = 2.0$ and $k_{unbind} = 25 \, s^{-1}$. As can be seen, the dwell time at small forces is mostly around $1 \, s$, whereas for the larger forces, the dwell time has a very wide distribution. This behavior is in a very good agreement with the experimental data (see Fig 4(a) in J.-D. Wen {\it et al.} \cite{Bustamante-2008}).

In this paper we have not addressed the process of frameshifting during the translation. When the ribosomal frameshifting is occured, the ribosome shifts reading frames mostly upstream by a single or more nucleotides along the mRNA and as a result, another protein will be produced. This process becomes more interesting when we see that many viruses benefit this possibility for producing their needed proteins from their single mRNA sequence \cite{Caliskan-2015}. Our model can be used in understanding possible scenarios in programming frameshifting.

We finally discuss the possible experimental investigations for checking our model for the translation process. An experiment could be achieved by doing the similar experiments using optical tweezers for more different sequences of mRNA. In the suggested experiment, the small subunit of the ribosome can be fixed, like Ref. \cite{Bustamante-2014}, and more stable mRNA sequences can be used. This experiment would be more useful in order to understand the details of programming frameshifting, as mentioned above.

In conclusion, we have shown that the results of our simple model for the translation process have a very good agreement with the experimental data in different situations. In this paper, it has been shown that the accommodation state of the ribosome plays a very important role in the translation process and it might be more important in the presence of the external force, which has not been considered very carefully in the previous models. 
As we discussed in the first part of the paper, there was a question regarding the mechanism of the translation process. Here we have shown that according to the experimental data and our results, the Brownian Ratchet mechanism can be the responsible mechanism for the ribosome. At the end we would like to emphasize that besides the effect of thermal fluctuations on unwinding dsmRNA, the ribosome contribution might be also considerable, depending on the values of $\Delta G$ and $k_{unbind}$.

\section*{Acknowledgments}

We thank R. Golestanian and L. Mollazadeh-Beidokhti for very helpful discussions. We thank Asal Atakhani and Maniya Maleki for very helpful comments on the manuscript.

\section*{Appendix A: Details of Translation Cycle}

\begin{figure}
\centering
\includegraphics[width=1\columnwidth]{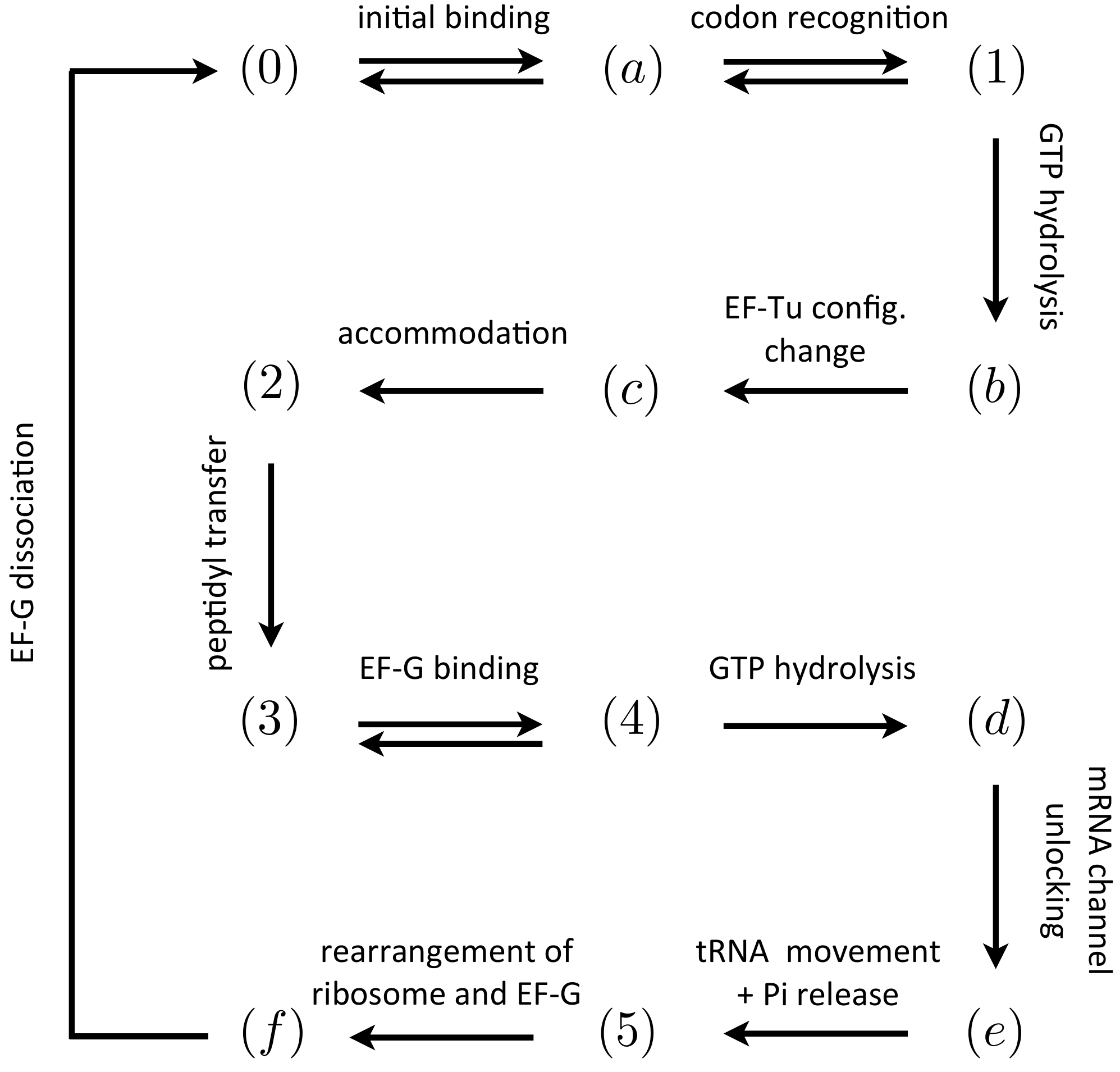}
\caption{Distinct steps of the ribosome translocation cycle. Here the numbers stand for the states that we have used in our model and the alphabets show intermediate states. 
Step $(0) \rightleftharpoons (a)$: a EF-Tu-dependent aa-tRNA binds to/unbinds from the ribosome. Step $(a) \rightleftharpoons (1)$: The mRNA codon is examined by the aa-tRNA anticodon. Step $(1) \rightarrow (b)$: GTP hydrolizes and therefore the incoming tRNA tightly binds to the mRNA codon in the A site of the ribosome. Step $(b) \rightarrow (c)$: The elongation 
factor of EF-Tu is deformed. Step $(c) \rightarrow (2)$: EF-Tu releases and the tRNA at the P site is joint to the tRNA at the A site through a peptidyl bond. Step $(2) \rightarrow (3)$: the polypeptide chain is transferred to the tRNA of the A site. Step 
$(3) \rightleftharpoons (4)$: the elongation factor of EF-G binds to the ribosome and promotes a ratchet like transition from the state (3) to state (4) and backward. Step $
(4) \rightarrow (d)$: GTP hydrolyzes and drives the unlocking of the mRNA channel, Step $(d) \rightarrow (e)$, and followed by the tRNA movement and releasing of Pi, Step $(e) \rightarrow (5)$. Step $(5) \rightarrow (f)$: rearrangement of the ribosome 
(re-locking) and elongation factor of EF-G are occurred. Step $(f) \rightarrow (0)$: 
EF-G dissociates and the ribosome goes to its initial state. }
\label{fig:App-Cycle}
\end{figure}

The translation process involves several states. In the model that has been proposed in this paper, we summarized these states in six main states as described in Fig. 2. Here we discuss the details of the translation cycle and show how one can derive the effective rates of our model. According to the experimental findings, the translation process is occurred in 13 states that have been shown in Fig. \ref{fig:App-Cycle} \cite{Frank-2010}. The intermediate sates that we have combined and considered them in our model implicitly, have been shown by alphabets of (a) to (e), and the states that we have already considered in our model have been shown by numbers of (0) to (5). One can find the details of the intermediate states in Ref. \cite{Frank-2010}. Using the experimental data for the ribosomal translation \cite{Pape-1998, Pape-1999,Stark-2002,Savelsbergh-2003,Wintermeyer-2004,Wilden-2006,Rodnina-2007}, we estimate the values of the transition rates for different states, which have been summarized in Table \ref{tab:exp_rate}.  
 
\begin{table}
\centering
\caption{The transition rate from state $i$ to state $j$, $k_{ij}$, for the translation cycle. The rates are estimated using the experimental data of references \cite{Pape-1998, Pape-1999,Stark-2002,Savelsbergh-2003,Wintermeyer-2004,Wilden-2006,Rodnina-2007}.}
\begin{ruledtabular}
\begin{tabular}{cccccccccccccccc}
parameter & $k_{0a}$ & $k_{a0}$ & $k_{a1}$ & $k_{1a}$  & $k_{1b}$ &$k_{bc}$ & $k_{c2}$ & $k_{23}$ \\
\hline rates (${\rm s^{-1}}$) & 110 & 25 & 100 & 0.2 & 250 & 60 & 3 & 50 \\
\hline parameter  & $k_{34}$ & $k_{43}$ & $k_{4d}$ & $k_{de}$ & $k_{e5}$ & $k_{5f}$ & $k_{f0}$\\
\hline rates (${\rm s^{-1}}$) & 150 & 140 & 250 & 35 & rapid & 5 & 20 \\
\end{tabular}
\end{ruledtabular}
\label{tab:exp_rate}
\end{table}

Using ``the concept of net rate constant'' \cite{cleland}, we can deduce the effect of the intermediate steps on the transition rates in our model. As an example we can find the rate of $\omega_{01}$ in terms of the rates of intermediate steps of $(0) \rightleftharpoons (a) \rightleftharpoons (b) \rightarrow (1)$ as 
\bea
\frac{1}{\omega_{01}} &=& \frac{1}{k_{0a}^{net}}+\frac{1}{k_{a1}^{net}},
\eea
where $k_{ij}$ denotes the rate of transition from state $i$ to state $j$, and $k_{0a}^{net}$ and $k_{a1}^{net}$ are  
\begin{subequations}
\bea
k_{0a}^{net} &=& k_{0a} \times \frac{k_{a1}^{net}}{k_{a0}+k_{a1}^{net}}, \\
k_{a1}^{net} &=& k_{a1} \times \frac{k_{1b}}{k_{1b}+k_{1a}}.
\eea
\end{subequations}

\section*{Appendix B: Analytical Derivation for the Translation Velocity}

\begin{figure}
\centering
\includegraphics[width=1\columnwidth]{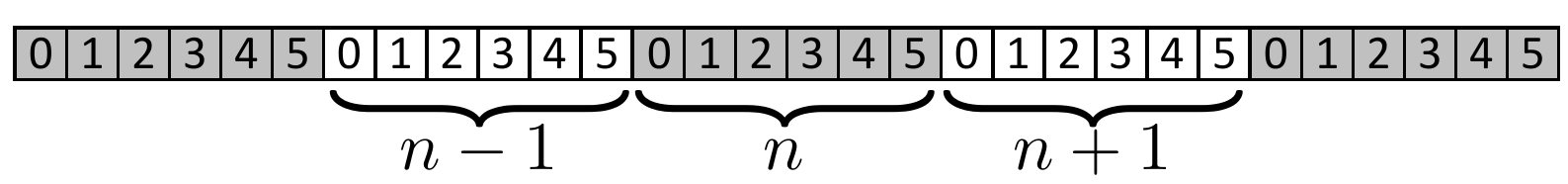}
\caption{The mean position of the ribosome is denoted by $n$, whereas its internal state is shown by $i$. }
\label{fig-lattice-model}
\end{figure}

To find the analytical description for the translation velocity, we consider a one-dimensional lattice; cf. Fig \ref{fig-lattice-model}, where the overall position of the ribosome is denoted by $n$ and the its internal state is denoted by $i$, as shown in Fig. \ref{fig-Model}.  The master equation governing this problem has been written in Eqs. (\ref{eq:master_a})-(\ref{eq:master_f}). After defining $P(n)$ as the probability for the ribosome to be in the position $n$, we can write the mean position as
\bea
\langle n \rangle = \sum_n n P(n) = \sum_n \sum_{i=0}^5 n p_i(n),
\eea
where $p_i(n)$ denotes the probability for the ribosome to be in the position $n$ and the internal state of $i$. By definition, the mean velocity of the ribosome can be written as
\bea
v \equiv \frac{\p}{\p t} \langle n \rangle = \sum_n \sum_{i=0}^5 n \frac{\p}{\p t} p_{i}(n).
\eea
A few simple calculations using the master equations of Eqs. (\ref{eq:master_a})-(\ref{eq:master_f}), lead to 
\bea
v = \omega_{50} \sum_n p_5(n).
\eea
In the steady state, we have $\p_t p_{i}(n) = 0$ and one can find all $p_i(n)$'s in terms of $p_5(n)$ as
\begin{subequations}
\bea
p_0(n) &=& \frac{\omega_{50}}{\omega_{01}} p_5(n), \\
p_1(n) &=& \frac{\omega_{50}}{\omega_{12}} p_5(n), \\
p_2(n) &=& \frac{\omega_{50}}{\omega_{23}} p_5(n), \\
p_3(n) &=& \frac{\omega_{50}}{\omega_{34}} \left( 1+ \frac{\omega_{43}}{\omega_{45}} \right)  p_5(n), \\
p_4(n) &=& \frac{\omega_{50}}{\omega_{45}} p_5(n).
\eea
\end{subequations}

Using the above equations and the condition of $\sum_n \sum_i p_i(n) = 1$, we have
\bea
\sum_n p_5(n) = \frac{1}{1+ \omega_{50} \left[ \frac{1}{\omega_{01}} + \frac{1}{\omega_{12}} + \frac{1}{\omega_{23}} + \frac{1}{\omega_{34}} \left( 1+ \frac{\omega_{43}}{\omega_{45}} \right) + \frac{1}{\omega_{45}} \right] }. \nonumber \\
\eea
Therefore we can find the mean velocity of the ribosome in terms of its internal rates as
\bea
v = \frac{1 }{ \frac{1}{\omega_{01}} + \frac{1}{\omega_{12}}  + \frac{1}{\omega_{23}} + \frac{1}{\omega_{34}} \left( 1 + \frac{\omega_{43}}{\omega_{45}} \right)  + \frac{1}{\omega_{45}} + \frac{1}{\omega_{50}} }. 
\eea
We note that since the distance between two neighboring sites is $1$ codon, the velocity is determined in terms of codon per second.

%
%
%

\section*{Appendix C: The algorithm of the simulation}

The six state model presented in this paper simulated using Gillespie algorithm, witch is a random selection method \cite{Gillespie}. This algorithm generates random reactive events consistent with master equations of the system. It has two random generating parts: (1) random selective reaction, and (1) random reaction time. In order to explain the details of the algorithm, we give an example regarding determination of value $P$, the fraction of actively opened to all opened base-pairs of dsRNA as depicted in Fig. \ref{fig:v-k_unbind}(b). 

We consider the ribosome is in the state (4) in the translational cycle as shown in Fig. 2, and it reaches a double stranded RNA. As discussed in the main text above, the local double strand should be somehow unwound for further translational process. This base-pair either can be broken passively by the thermal fluctuations or actively by the ribosome. As mentioned in the main text, the rate of the passive process is denoted by $k_{unbind}$, and the rate of the active process is shown by $\omega_{45}$, Eq. (\ref{eq:omega45}). Now a random number $0 \leq \zeta_1 <1$ is drawn, and depending on its value the next decision is taken as
\begin{eqnarray}
{\rm If} \; 0 \leq \zeta_1 < \frac{k_{unbind}}{k_{unbind} + \omega_{45}} \; &:& \; {\rm passive \; unwinding}, \nonumber \\
{\rm otherwise} \; \hspace{2.5cm }&:& \; {\rm active \; unwinding}. \nonumber
\end{eqnarray}
This process is happened in the time interval of $\Delta t$ as
\bea
\Delta t = \frac{-1}{k_{unbind} + \omega_{45}} \times \ln (1 - \zeta_2 ), \nonumber
\eea
where $  0 \leq \zeta_2 < 1$ is another random number that is using in the Gillespie algorithm. During the translation process, we count the number of base pairs that are broken passively and actively. At the end we can determine the value of $P$ as 
\bea
P = \frac{{\rm Number ~ of ~ actively ~ broken ~ base ~ pairs}}{{\rm Total ~ number}}. \nonumber
\eea
This quantity has been shown in Fig. \ref{fig:v-k_unbind}(b) for different values of $k_{unbind}$.


\begin{thebibliography}{}

\bibitem{Cell}
B. Alberts {\it et al.}, {\it Molecular Biology of the Cell} (Garland, New York, 2007), 5th ed..

\bibitem{Ramakrishnan-2009}
T.M. Schmeing, and V. Ramakrishnan, Nature (London) {\bf 461}, 1234 (2009).

\bibitem{Yusupov-2001}
M.M. Yusupov {\it et al.} Science {\bf 292}, 883 (2001).

\bibitem{Harms-2001}
J. Harms {\it et al.} Cell {\bf 107}, 679 (2001).

\bibitem{Schuwirth-2005}
B. S. Schuwirth,{\it et al.} Science {\bf 310}, 827 (2005).

\bibitem{Selmer-2006}
M. Selmer {\it et al.}, Science {\bf 313}, 1935 (2006).

\bibitem{Ishida-2014}
H. Ishida and A. Matsumoto, PLoS ONE {\bf 9}, e101951 (2014).

\bibitem{Garai-2009}
A. Garai {\it et al.} Phys. Rev. E {\bf 80}, 011908 (2009).

\bibitem{PingXie-2013}
P. Xie, PLoS ONE {\bf 8}, e70789 (2013); Eur. Biophys. J. {\bf 42}, 347 (2013).

\bibitem{Bailey-2014}
B. L. Bailey, K. Visscher, and J. Watkins, Phys. Biol. {\bf 11}, 016009 (2014).



\bibitem{Chacon-2003}
P. Chac\'{o}n, F. Tama, and W. Wriggers, J. Mol. Biol. {\bf 326}, 485 (2003).


\bibitem{Stember-2009}
J.N. Stember, and W. Wriggers, J. Chem. Phys. {\bf 131}, 074112 (2009)


\bibitem{Bustamante-2008}
J.-D. Wen {\it et al.}, Nature {\bf 452}, 598 (2008).


\bibitem{Bustamante-2015}
S. Yan {\it et al.}, Cell {\bf 160}, 870 (2015).



\bibitem{Wang-2002}
H. Wang, and G. Oster, Appl. Phys. A {\bf 75}, 315 (2002).

\bibitem{Bustamante-2011}
X. Qu {\it et al.} Nature (London) {\bf 475}, 118 (2011).


\bibitem{Wintermeyer-2004}
W. Wintermeyer, {\it et al.} Biochem. Soc. Trans. {\bf 32}, 733 (2004).

\bibitem{Frank-2010}
J. Frank, and R. L. Gonzalez Jr. Annu. Rev. Biochem. {\bf 79}, 381 (2010).


\bibitem{Blanchard-2004}
S.C. Blanchard, {\it et al.} Nat. Struct. Mol. Biol. {\bf 11}, 1008 (2004).

\bibitem{Daviter-2006}
T. Daviter, K.B. Gromadski, M.V. Rodnina, Biochimie {\bf 88}, 1001 (2006).


\bibitem{Noller-2002}
H. F. Noller {\it et al.}, FEBS Lett. {\bf 514}, 11 (2002). 

\bibitem{Voorhees-2009}
R.M. Voorhees {\it et al.}, Nat. Struct. Mol. Biol. {\bf 16}, 528 (2009).

\bibitem{Agirrezabala-2008}
X. Agirrezabala {\it et al.},  Mol. Cell {\bf 32}, 190 (2008).

\bibitem{Valle-2003}
M. Valle {\it et al.}, Cell {\bf 114}, 123 (2003).

\bibitem{Howard}
J. Howard, {\it Mechanics of Motor Proteins and the Cytoskeleton} (Sinauer Associates, Sunderland, MA, 2001).

\bibitem{Rob-Philips}
R. Phillips, J. Kondev, and J. Theriot, {\it Physical biology of the cell} (Garland Science, Taylor $\&$ Francis, New York, NY, 2009).

\bibitem{Gillespie}
D.T. Gillespie, J. Phys. Chem. {\bf 81} 2340 (1977).


\bibitem{Bustamante-2014}
T. Liu {\it et al.} eLife {\bf 3}, e03406 (2014).


\bibitem{Seyedtaghi-2005}
S. Takyar, R. P. Hickerson, H. F. Noller, Cell {\bf 120}, 49 (2005). 




\bibitem{Caliskan-2015}
N. Caliskan, F. Peske, and M.V. Rodnina, Trends Biochem. Sci. {\bf 40}, 265 (2015). 



\bibitem{Stember-2007}
J.N. Stember, and G.S. Ezra, Chem. Phys. {\bf 337}, 11 (2007).


\bibitem{cleland}
W.W. Cleland, Biochem. {\bf 14}, 3220 (1975).



\bibitem{Pape-1998}
T. Pape, W. Wintermeyer, M.V. Rodnina, EMBO J. {\bf 17}, 7490 (1998).


\bibitem{Pape-1999}
T. Pape, W. Wintermeyer, M.V. Rodnina, EMBO J. {\bf 18}, 3800 (1999).


\bibitem{Stark-2002}
H. Stark {\it et al.}, Nature Struct. Mol. Biol. {\bf 9}, 849 (2002).




\bibitem{Savelsbergh-2003}
A. Savelsbergh {\it et al.}, Mol. Cell {\bf 11}, 1517 (2003).



\bibitem{Wilden-2006}
B. Wilden, A. Savelsbergh, M.V. Rodnina, W. Wintermeyer, Proc. Natl. Acad. Sci. U.S.A. {\bf 103}, 13670 (2006).


\bibitem{Rodnina-2007}
M.V. Rodnina, M. Beringer, and W. Wintermeyer, Trends Biochem. Sci. {\bf 32}, 20 (2007).



\bibitem{Zhang-2001}
Y. Zhang, H. Zhou, Z.-C. Ou-Yang, Biophys. J. {\bf 81},  1133 (2001).

\end{thebibliography}
\end{document}